\documentclass[manuscript]{aastex701}

\begin{document}

\title{A Pre-Main Sequence Binary Viewed Through its Circumbinary Disk: A New Phase for KH 15D}

\author[orcid=0000-0001-5924-3531,gname=Catrina,sname=Hamilton-Drager]{Catrina M. Hamilton}
\affiliation{Dickinson College, Department of Physics and Astronomy}
\email[show]{hamiltoc@dickinson.edu}

\author[gname=Adi,sname=Chacko]{Adi B. Chacko}
\affiliation{Dickinson College, Department of Physics and Astronomy}
\email{chackoa@dickinson.edu}

\author[orcid=0000-0002-3485-8924,gname=George,sname=Carson]{George A. Carson} 
\affiliation{Tufts University, Department of Physics and Astronomy}
\email{George.Carson.@tufts.edu}

\author[orcid=0000-0002-1655-6041,gname=Aidan,sname=Pidgeon]{Aidan J. Pidgeon}
\affiliation{Space Telescope Science Institute}
\email{ajpidgeon@gmail.com}

\author[orcid=0009-0006-0731-7192,gname=Catherine Nicole,sname=Foley]{C. Nicole Foley}
\affiliation{Dickinson College, Department of Physics and Astronomy}
\email{foleyca@dickinson.edu}

\author[gname=Hillel,sname=Finder]{Hillel Finder}
\affiliation{Dickinson College, Department of Physics and Astronomy}
\email{hftwin@gmail.com}

\author[gname=Gimmy,sname=Kyle]{Kyle Gimmy}
\affiliation{Rensselaer Polytechnic Institute, Department of Physics, Applied Physics, and Astronomy}
\email{gimmyk@rpi.edu}

\author[orcid=0000-0003-1378-1865,gname=Aleezah,sname=Ali]{Aleezah Ali}
\affiliation{Wesleyan University, Department of Astronomy}
\email{aleezaha@gmail.com}

\author[orcid=0000-0001-7624-3322,gname=William,sname=Herbst]{William Herbst}
\affiliation{Wesleyan University, Department of Astronomy}
\email{wherbst@wesleyan.edu}

\author[orcid=0000-0003-4734-3345,gname=Dirk,sname=Froebrich]{Dirk Froebrich}
\affiliation{University of Kent, Centre for Astrophysics and Planetary Science, School of Engineering, Mathematics and Physics}
\email{df@kent.ac.uk}

\author[orcid=0000-0001-5306-6220,gname=Brian,sname=Skiff]{Brian A. Skiff}
\affiliation{Lowell Observatory}
\email{bas@lowell.edu}

\author[orcid=0009-0002-1717-419X,gname=Bart,sname=Staels]{Bart Staels}
\affiliation{American Association of Variable Star Observers}
\email{staels.bart.bvba@telenet.be}

\author[orcid=0009-0002-1717-419X,gname=Shawn,sname=Dvorak]{Shawn Dvorak}
\affiliation{American Association of Variable Star Observers}
\email{sdvorak@rollinghillsobs.org}

\author[orcid=0000-0001-6496-0252,gname=Jochen,sname=Eisl\"offel]{Jochen Eisl\"offel}
\affiliation{Th\"uringer Landessternwarte, Sternwarte 5, D-07778 Tautenburg, Germany}
\email{jochen@tls-tautenburg.de}

\author[gname=Teryn,sname=Huff]{Teryn Huff}
\affiliation{Dickinson College, Department of Physics and Astronomy}
\email{hufft@dickinson.edu}

\author[gname=Gavin,sname=Frueh]{Gavin Frueh}
\affiliation{University of Iowa, Department of Physics and Astronomy}
\email{gavin-frueh@uiowa.edu}

\author[gname=Abigail,sname=Mead]{Abby Mead}
\affiliation{Dickinson College, Department of Physics and Astronomy}
\email{abbymead37@gmail.com}

\author[gname=Lia,sname=Gilmore]{Lia M. Gilmore}
\affiliation{University of Connecticut, Physics Department}
\email{lia.gilmore@uconn.edu}

\author[gname=Robert,sname=Boyle]{Robert J. Boyle}
\affiliation{Dickinson College, Department of Physics and Astronomy}
\email{boyle@dickinson.edu}

\begin{abstract}
The binary T Tauri system, KH 15D, is poised to provide unprecedented detail of gas and ices located between $\sim$3-5 AU of its circumbinary disk over the next few years. We analyze for the first time a complete set of ground-based time series photometry of the KH 15D system for which the 2022 October - 2025 April data have never been presented. By combining the latest $I$-band data with historical measurements, we show that the era of large amplitude photometric variability has ended. We define a revised photometric period of $P$ = 48.365 $\pm$ 0.005 days, which accurately phases data over a time span of seventy-five years. We adopt magnitudes and colors for the stars that define the epochs when either star A (1995-2008) or star B (2010-2012) is the primary source of the out-of-eclipse brightness of the system. We determine for the first time the effective optical depth for the system between 2002 and 2025 assuming gray extinction. The current out-of-eclipse system color is consistent with a combination of the light from both stars, supporting the previously suggested model in which the trailing edge of the circumbinary disk is more transparent than the leading edge. We continue to see evidence for ``clumps'' and variable transparency along ingress and egress when the stars are probing the occulting material and suggest that observations made during these phases should permit studies of the physical and chemical nature of the planet-forming zone with the circumbinary disk. 

\end{abstract}

\section{Introduction}
\label{sec:intro}

\begin{figure*}[htbp]
\centering
\includegraphics[scale=0.7]{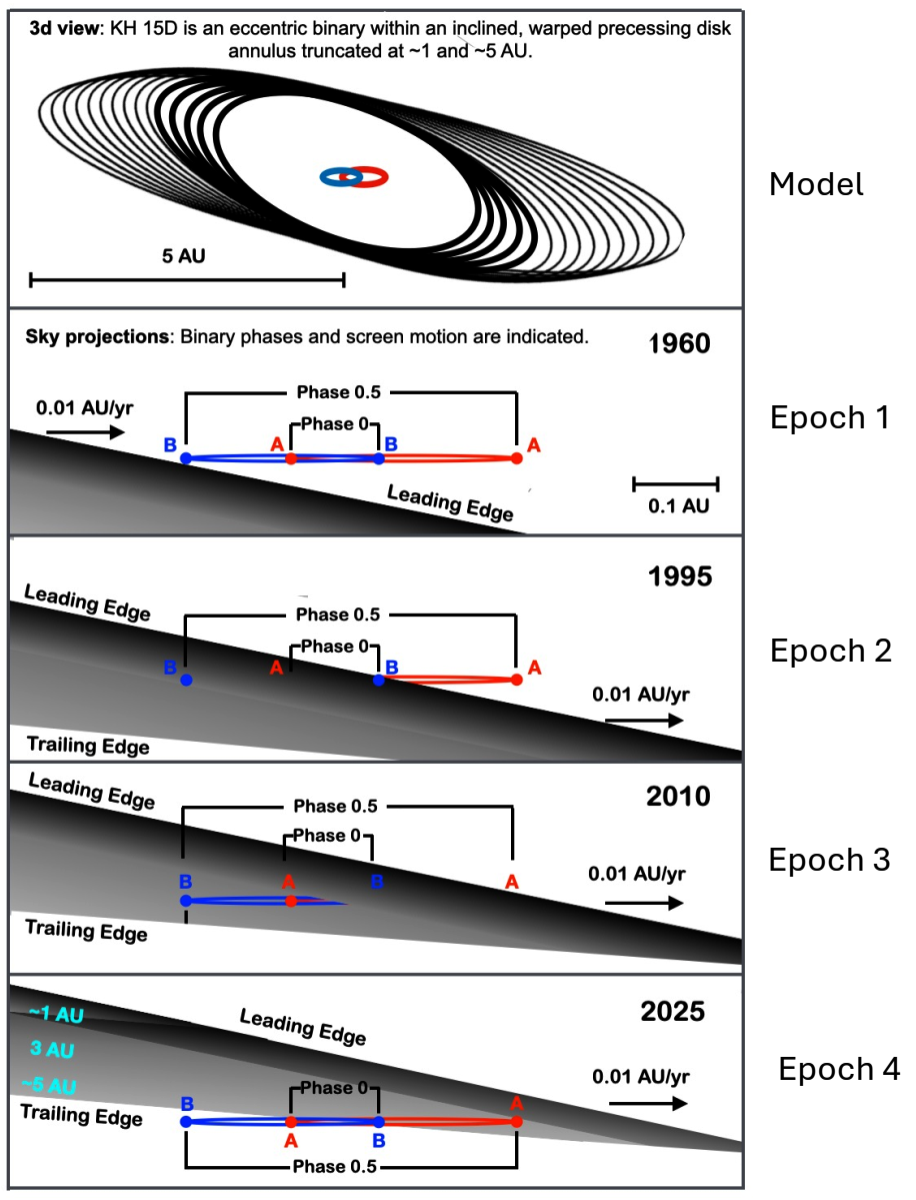}
\caption{A schematic representation of the KH 15D system. The top panel shows a 3D view adapted from \citet{Winn04} and \citet{Poon21}. The binary orbits are shown in red (star A) and blue (star B).} The binary is viewed nearly edge-on (i = 89$^\circ$). Sky projections are shown in the bottom panels for four epochs. The warped disk projects as a screen that moves from left to right across the panels as it precesses, covering different parts of the stellar orbits. The photometric period is 48.365 days and the locations of the stars at what we define as phase 0, minimum projected separation near periastron, and phase 0.5, maximum projected separation near apastron, are labeled. As discussed in the text, the leading edge is identified as the inner edge of the disk at $\sim$1 AU and appears sharper, more opaque and more highly inclined than the trailing edge at $\sim$5 AU, which is fuzzier, less inclined and more transparent.
\label{fig:schematic}
\end{figure*}

V582 Mon was known only as an irregular variable in NGC 2264 \citep{Badalian70, Kukarkin72} until 1995, when its large-amplitude periodic photometric behavior was serendipitously discovered  by \citet{KearnsHerbst98}. KH 15D, its common monikor since that paper, is now recognized as a pre-main sequence binary embedded in a circumbinary (CB) disk that is truncated at $\sim$1 AU by the action of the binary and at $\sim$5 AU, possibly by a protoplanet. As illustrated in the top panel of Figure~\ref{fig:schematic} (the Model), the CB disk, represented by concentric ellipses, is inclined to the orbital plane and warped such that it precesses as a unit on a time scale of $\sim$10$^3$ years \citep{ChiangMurrayClay04, Winn04, Winn06, Poon21}. The thicker rings pictured in the model represent the warped portion of the CB disk. The period of the binary is $\sim$48.6 days and its eccentric orbit is viewed nearly edge-on \citep{Johnson04, Poon21}. Since $\sim$1960, due to precession, the CB disk has been oriented such that it covers various portions of the binary orbit as seen from Earth. In the bottom four panels of Figure~\ref{fig:schematic}, we illustrate how precession has gradually moved the disk across the orbit resulting in eclipse-like light curves that evolve with time. Epochs 1-4 identify periods of time that are characterized by specific light curve variability. In Epoch 1 (1960), the entire binary is visible, and the system is at its maximum brightness. The observed photometric variations during this time are due to the disappearance of star B behind the leading edge of the warp in the CB disk. In Epoch 2 (1995), the leading edge of the CB disk has advanced such that nearly all of the orbit of star B is obscured. The system's maximum brightness is now characterized by the brightness of star A alone. The deep eclipses seen during this time period happen as star A now disappears behind the leading edge of the CB disk. A characteristic feature of the observed light curve during this time was a reversal in brightness near mid-eclipse due to the approach of star B close to the edge of the occulting disk. This feature has been used to identify the photometric period of the system. In Epoch 3 (2010), precession has now carried the warp in the CB disk across the entire binary orbit. Fluctuations in brightness are still seen, presumably by reflected light and/or transmitted light through the disk. In Epoch 4 (2025), the warp has precessed such that the orbit of star B is now being revealed by the trailing edge. These time periods are discussed in further detail in Section~\ref{sec:photperiod}. As is evident from the figure and the recent data discussed here, the era of large amplitude photometric variability caused by the opaqueness of the leading edge of the disk has ended and we have entered a new phase for the system, characterized by much lower amplitude photometric variability, as the binary is now seen through the apparently semi-transparent trailing edge of its disk.

Achieving this basic understanding of the system has involved dozens of observational, theoretical and modeling efforts over nearly three decades, including photographic photometry from plate archives that go back to $\sim$1950 \citep{Winn03, JohnsonWinn04, Johnson05, Maffei05}, ground-based $VRIJHK$ photometry from 1995 to present (Table 3 and references therein), $gz$ photometry from 2018 to present \citep{Lamitina25}, space-based studies at X-ray, optical and infrared wavelengths \citep{Deming04, HerbstMoran05, Arulanantham16, Arulanantham17}, radio studies \citep{Aronow18}, ground-based infrared and high resolution spectroscopic studies \citep{Hamilton03, Tokunaga04, Johnson04, Lawler10, Mundt10, Hamilton12} and theory and modeling work \citep{Agol04, Winn04,ChiangMurrayClay04,Winn06,SilviaAgol08,LodatoFacchini13,Smallwood19,Fang19, GarciaSoto20,Poon21}.

If not occulted by its disk, KH 15D would be known as an ordinary Class 3, weak T Tauri (WTTS) binary member of NGC 2264, at a distance of 780 pc, with an inferred age of $\sim$2 Myr and no disk signature in the infrared or radio \citep{Hamilton01, Aronow18, GarciaSoto20}. Its total mass is 1.3 M$_\odot$ and the mass ratio of the components is $\sim$0.86 \citep{Poon21} so the system is not dissimilar to the Solar System, {\it albeit} with a binary at the center. Within the last fifteen years, with the help of the Kepler space telescope as well as the Transiting Exoplanet Survey Satellite (TESS), binary star systems have been observed to host planets \citep{Doyle11, Welsh12, Kostov20}. These systems have shown us that binary systems can host stable CB orbits for planets and KH 15D's CB disk is, therefore, a plausible site of on-going planet formation.  

An important feature of the projected CB disk that we illustrate in Figure~\ref{fig:schematic} is that its leading edge is nearly opaque and can be well-modeled as a ``knife edge", but its trailing edge seems not as sharply defined and more transparent \citep{GarciaSoto20}. \citet{Poon21} identify the sharp leading edge with the inner edge of the disk at $\sim$1 AU, and the fuzzier trailing edge with its outer edge at $\sim$5 AU, as indicated in Epoch 4 of Figure~\ref{fig:schematic}. The apparent transition from a nearly opaque leading (inner) edge to a more transparent trailing (outer) edge means that we have an enhanced opportunity now and in the years to come to probe the physical and chemical state of the disk. Interestingly, the age of the KH 15D disk, $\sim$2 Myr, and location probed, $\sim$3-5 AU from the central binary, correspond to the time and site inferred for chondrule formation in the Solar System \citep{Russell18}.    

KH 15D's optical spectrum reveals forbidden emission lines ([NII] and [OIII]) and high velocity wings on its H$\alpha$ line, indications of active accretion and outflow \citep{Hamilton03, Mundt10,Hamilton12}; however, see also \citet{Fang19} for another interpretation. The binary may be the driving source of a bipolar jet seen in the light of shocked H$_2$ and CO, although there is a curious offset of the jet center from its photocenter that amounts to $\sim$1 arc-sec or $\sim$800 AU \citep{Tokunaga04, Aronow18}. KH 15D is definitely a source, therefore, in which gas flow in an accreting CB system can be monitored, which has astrophysical relevance on both stellar and galactic scales \citep{Artymowicz91, Valli24}. 

In this contribution, we present photometric observations made over seven seasons from 2018 October through 2025 April. Our results, now based on more than seventy five years of observations, clearly show the success of the models that have been in place since 2004 and expanded upon in recent years, but also point out their limitations. In particular, as illustrated in Figure~\ref{fig:schematic}, it now appears to be the case that the CB disk has an optical depth gradient across it and that light from both stars has likely been transmitted through the ring all along. In Sections 2 and 3 we describe the observations that contributed to our current observing campaign and our results. In Section 4 we discuss the implications of our results and suggestions for future observations of KH 15D and in Section 5 we present our conclusions. 

\section{Observations, Data Reduction, and Photometry}
\label{sec:obs}
Multiple observatories provided data on KH 15D for the observing campaign that ran between 2018 October and 2025 April. Table~\ref{tab:obs} lists the observatories, location, telescope aperture size, number of observations contributed, filters used, and the seasons in which observations were made. 

In the following sections, we describe the equipment and observing procedures at each observatory, as well as the photometric process that produced the data used in this analysis.

\begin{deluxetable}{lcccccc}[h!]
\tablecolumns{3}
\tabletypesize{\footnotesize}
\tablecaption{Observatories Contributing to the 2018-2025 Campaign}
\label{tab:obs}
\tablehead
{ 
\colhead{Observatory} & \colhead{Location} & \colhead{Aperture Size} & \colhead{Filters} & \colhead{Number of Observations} & \colhead{Seasons}  }
\startdata
Cerro Tololo Inter-American Observatory (CTIO) & Chile & 1.3 m & $VRIJHK$ & 163 & 2018-2019 \\ 
Lowell Observatory (LOWL) & USA (AZ) & 0.7 m/1.1 m & $VIz$ & 307 & 2020-2025 \\ 
Michael L. Britton Observatory (BRIT) & USA (PA) & 0.6 m & $VRI$ & 213 & 2020-2025 \\
The Liverpool Telescope (LT) & Canary Islands & 2.0 m & $BVuriz$ & 87 &  2020-2022 \\ 
Telescope Altiplano de Granada (TAGRA) & Spain & 0.5 m & $VRIz$ & 175 & 2022-2025 \\ 
Rolling Hills Observatory (ROLL) & USA (FL) & 0.4 m & $VI$ & 130 & 2022-2025 \\ 
\enddata
\vspace{-9mm}
\end{deluxetable} 

\subsection{Cerro Tololo Inter-American Observatory (CTIO)}
KH 15D was observed over many nights during the 2018-2019 observing season with the CTIO 1.3-m telescope operated by the Small and Moderate Aperture Research Telescope System. These data were taken in the $VRIJHK$ bands using A Novel Dual Imaging CAMera (ANDICAM; \citet{DePoy03}) before the system was retired in 2019 July. Each optical observation included four 150s exposures. All images have a 10\arcmin.2 × 10\arcmin.2 field of view. These data were reduced and photometered using similar procedures described in \citet{WindemuthHerbst14}, \citet{Arulanantham16}, and \citet{Aronow18}. 

\subsection{Michael L. Britton Observatory}
Following the end of the KH 15D coverage by ANDICAM, coordinated efforts to observe the system with the 0.6-m (24-inch) Britton telescope at Dickinson College commenced. The telescope is a Ritchey-Chr\'etien (RC) telescope, which is equipped with an SBIG STXL-6303E 3072×2048 CCD, resulting in a field of view of 15\arcmin.9 x 10\arcmin.6. Sets of five 60-second exposures were obtained, with priority given to observations in the Bessell $I$ band. Images were bias-subtracted, dark-subtracted, and flat-fielded with the Data Processor module of AstroImageJ (AIJ; \citet{Collins17}). Images in each set of five exposures were aligned using the AIJ Stack Aligner tool, then average combined into a single image to increase the target's signal-to-noise ratio.

Aperture photometry was performed on the average combined images using AIJ. The aperture radius was chosen to be 1.5 times the average FWHM of KH 15D over the course of the night, with a gap of five pixels (1.5\arcsec) and a sky annulus width of five pixels (1.5\arcsec). This minimized the incursion of light from the nearby bright star HD 47887. We used the same photometric comparison stars as \citet{GarciaSoto20}, labeled stars C and F in \citet{Hamilton05} (referred to as stars 02 and 01 in \citet{GarciaSoto20}, respectively, and stars 130 and 139 by \citet{Flaccomio99}). We photometered star E (see \citet{Hamilton05}) to determine the stability of the comparison stars in each season. In each case, the variability was less than 0.01 magnitudes in all filters used. 

\subsection{Lowell Observatory} 

\subsubsection{0.7-m (31-inch) Telescope on Anderson Mesa}
The Lowell Observatory 0.7-m telescope was used to obtain sets of 5 x 60s exposures through standard Bessell $V$ and $I$ filters manufactured by Omega Optical. The telescope was equipped with a 2K×2K Loral CCD. The camera employed a 2:1 focal reducer, giving a plate-scale of 0\arcsec.91/pixel covering a field of about 15\arcmin x 15\arcmin. At least ten twilight flat-field images were taken during cloud-free dusk/dawns along with batches of usually 20 bias frames. Flats from adjacent nights were used as necessary. Conventional bias and flat-correction was done using in-house IDL scripts. Differential photometry was performed using the commercial software `MPO Canopus'\footnote{https://minplanobs.org/BdwPub/php/displayhome.php}. We adopted magnitudes for the comparison stars C, F, and E from \citet{Hamilton05}. Because of scattered light from nearby HD 47887, the measuring apertures were restricted to 13 pixels diameter and 'sky' annulus 5 pixels wide separated by 5 pixels from the star apertures. Nightly rms scatter in the comparison stars was 0.006-0.008 mag. This telescope was closed in 2020 July.

\subsubsection{1.1-m (42-inch Hall) Telescope on Anderson Mesa}
The majority of the Lowell 1.1-m data were obtained in 3x3-binning mode (1\arcsec.1/pixel), where the field covers about 24\arcmin x24\arcmin. The data were reduced in the same way as the 0.7-m data. Measuring apertures were fixed at 7 pixels in diameter for the stars, with narrow `sky' annuli 5 pixels wide separated by 3 pixels from the star aperture. Sloan $z$ magnitudes were adopted from ATLAS `refcat2' (\citet{Tonry18}). The nightly scatter in the comparison stars is similar to the 0.7-m data.

\subsection{Liverpool Telescope (LT)}
One image each in the Sloan Digital Sky Survey (SDSS) $u$, $r$, $i$ filters, and Johnson $B$ and $V$ bands was obtained on clear nights beginning in late September 2020, with the SDSS $z$ filter being added shortly thereafter. These data were automatically reduced by the Liverpool Telescope's IO:O camera pipeline (\citet{Barnsley16}). In 2021-2022, the decision was made to monitor KH 15D in only the Johnson $B$ and $V$ and SDSS $i$ filters. For this analysis, we have only examined the $V$, $r$, and $i$ images from 2020-2021  (program number PL20B07), and the $V$ and $i$ images from 2021-2022  (program number XPL21B07).

Photometry of the LT data in $i$ demonstrated that the comparison stars C and F were in some cases saturated and in others, near the linearity limit. As such, we opted to use the ``Auto comparison stars" method in AIJ, which selects stars of similar brightness to the target for differential photometry. This process selected four stars that are known T Tauri stars. Photometry of each of these stars was performed using the other three as comparisons to assess their variability throughout the observing season. Two were found to be less variable than the others (on the order of $\pm0.02$ mag in $i$) and were chosen to be our comparison and check stars, respectively. Information about these stars is listed in Table~\ref{tab:compstars}.

\begin{deluxetable}{lccccc}[htbp]
\tablecolumns{5}
\tablecaption{Comparison and Check Stars for the LT Data{$^a$}}
\tablehead
{ 
\colhead{Star} & \colhead{RA (J2000)} &  \colhead{Dec (J2000)} & \colhead{$r$} & \colhead{$i$} 
}
\startdata
Mon-000186 & 100.35958 & 9.58013 & 16.001 & 15.447 \\ 
Mon-000134 & 100.31010 & 9.44952 & 16.816 & 15.343 \\ 
\enddata

\tablecomments{
\newline
$^a$ These values come from \citet{Venuti14}.
}
\label{tab:compstars}
\vspace{-9mm}
\end{deluxetable} 

Photometry was performed using the ``Auto Variable Apertures" option in AIJ with the FWHM factor set to 1.5. Due to the brightness of the nearby star HD 47887, our inner background annulus was set to extend two pixels beyond the aperture radius. The width of the background annulus was set to 5 pixels. To transform these data to Bessell $I$, we utilized information from \citet{Jordi06} and derived the following equation.

\begin{equation}
    \label{eq:transform}
    I = -(0.247 \pm0.003) \frac{(r-i)}{1.007 \pm 0.005} + i
\end{equation}

Since we had no direct $r$ measurements during the 2021-22 season, we calculated the average of the difference between the transformed $I$ and $i$ magnitudes out of eclipse for the 2020-21 season. This offset was then applied to the $i$ data obtained during 2021-22. When plotting all the $I$-band data, we noted that the LT data appeared to be offset from the others. We compared the LT data to data that which was obtained on the same nights at the Lowell Observatory and calculated an average offset. This offset was added to the two seasons. Since an offset needed to be applied, we do not use the LT data in any formal calculations regarding color for the system, and we do not include the LT data in Table~\ref{tab:photometry}, which contains all the new data presented in this paper.

\subsection{Telescope Altiplano de Granada (TAGRA)}
Observations began in 2022 November at TAGRA, located in the south of Spain at the Pixelskies Observatory\footnote{https://www.pixelskiesastro.com/}. The telescope is a 0.5-m (20-inch) $f$/7.7 Planewave telescope and observations of KH 15D are made with Bessell $V$, $R$, and $I$ filters, as well as the Sloan $z$ filter. The camera used is a CMOS ZWO ASI 2600MM Pro Mono camera, 6248 x 4176, with 3.76 x 3.76 micron pixels. Exposure times were typically 240 s in $V$, $R$, and $I$, and 300 s in $z$. The data are reduced using MaxIm DL and aperture photometry is performed with the AAVSO VPhot software. The aperture radius was set to 1.5 times the FWHM for each image.

\subsection{Data Collected Through the HOYS Program}
In collaboration with the citizen science program HOYS \citep{Froebrich18}, we received over 8,000 processed images taken by various observers at a number of different observatories. For this paper, we focused solely on observations that were made in the $I$ band. As such, we were able to extract useful photometry from only two observatories: Rolling Hills Observatory, and Thuringian State Observatory (denoted as ROLL and TLS in Table~\ref{tab:photometry}). Rejected images were generally plagued by poor seeing and/or tracking/trailing issues, or photometry that produced very large errors. Photometry was performed on what were deemed ``good" images using the same AIJ methods described in the section on the Michael L. Britton Observatory.

\subsection{Comparison of Photometric Observations}
\label{sec:comparison}
\begin{figure*}[htbp]
\plottwo{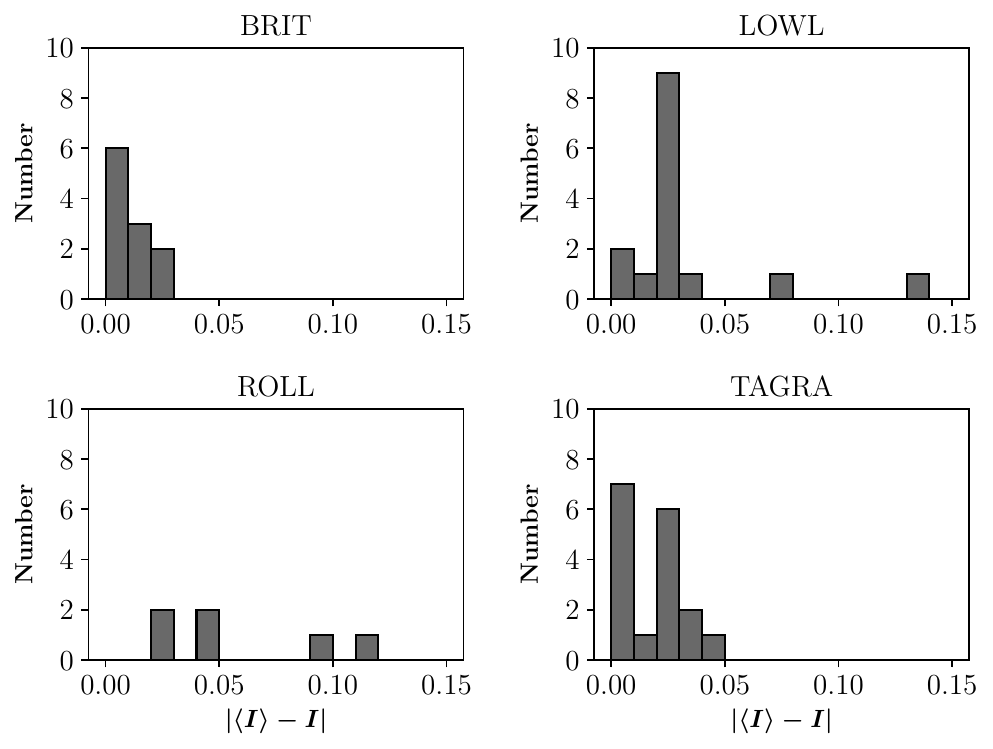}{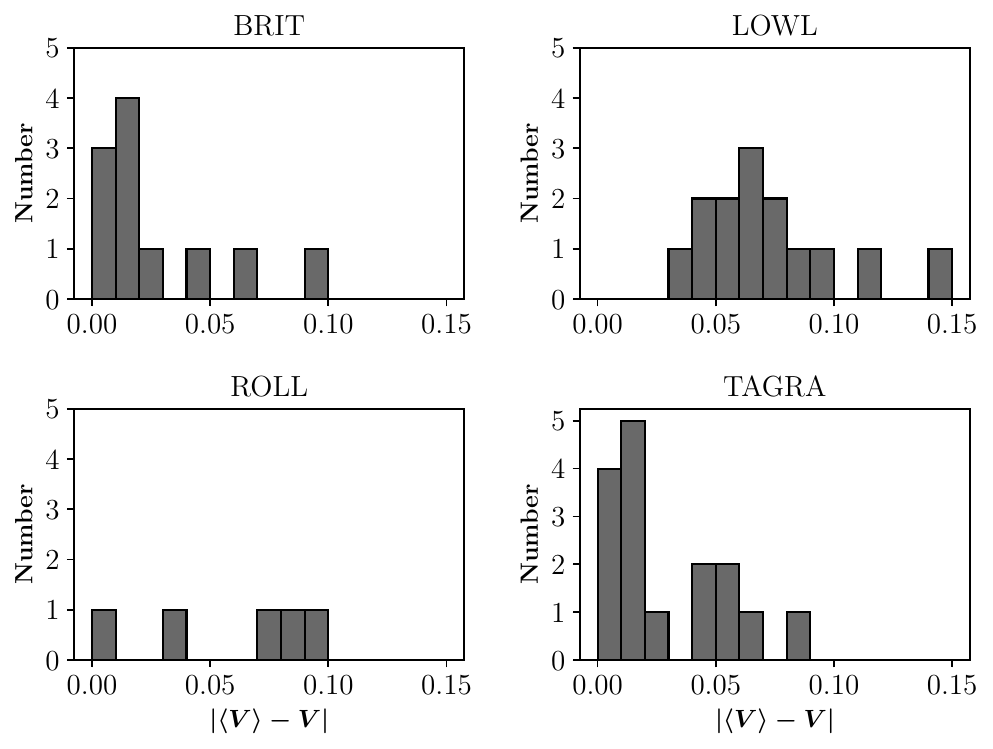}
\caption{The absolute difference between each observatory's measured $I$ or $V$ band magnitude from the weighted mean magnitude for each night out of eclipse. On average, the datasets are in agreement to $\sim$ 0.05 in $I$ and $\sim$ 0.1 in $V$. There is a systematic offset in $V$ with the data from Lowell Observatory, but it remains within the 1$\sigma$ uncertainty.}
\label{fig:VIhist}
\end{figure*}

To assess our ability to combine these different datasets, we compared measurements obtained on the same night at different observatories outside of eclipse. ``Out-of-eclipse" was set to be between phases 0.4 and 0.6. Phase = 0.0 represents mid-eclipse. Some individual points were in excellent agreement, while others displayed differences as large as $\sim$ 0.1 mag on a single night. Both star A and B are T Tauri stars, which are intrinsically variable at the 0.1 mag level due to spot patterns that can change on time scales of months to years. Additionally, the rotational modulation of these spots can affect the overall observed brightness. Disentangling that effect from systematic errors between observatories and small extinction changes present due to dust not confined precisely to the circumbinary ring plane is becoming more of a challenge. On the nights where we had multiple measurements from different observatories, we calculated an average weighted flux to determine the mean magnitude of the system for the night. We then measured the difference between each observatory's $I$ magnitude and the nightly mean. The results of this analysis are shown in Figure~\ref{fig:VIhist}, left. The data sets are consistent with each other to within $\sim$ 0.05 mag. We performed the same analysis in $V$ and found that there is a larger difference of about $\sim$ 0.1 mag among the datasets in this filter (see Figure~\ref{fig:VIhist}, right). This is likely due to the fact that the system is fainter in $V$ and the majority of the telescopes used in this study range in aperture size from 0.4 m - 0.6 m. We also find a systematic offset in $V$ relative to the Lowell Observatory data. This discrepancy is likely due to slight differences in the $V$ filters and remains within the 1$\sigma$ uncertainty. Table~\ref{tab:photometry} contains the new data appended to the data set of \citet{GarciaSoto20} representing a compilation of all the modern-day CCD observations.

\begin{deluxetable*}{ccccccccccccccc}[htbp]
\tablecolumns{14}
\tablecaption{\textit{VRIJHK} Photometry of KH 15D}
\tablehead
{ 
\colhead{Julian Date} &  \colhead{\textit{V}}  & \colhead{$\sigma_V$}  & \colhead{\textit{R}}  & \colhead{$\sigma_R$} &  \colhead{\textit{I}}  & \colhead{$\sigma_I$} &  \colhead{\textit{J}} & \colhead{$\sigma_J$} & \colhead{\textit{H}} & \colhead{$\sigma_H$} & \colhead{\textit{K}}  & \colhead{$\sigma_K$} &\colhead{Obs.}  &\colhead{Ref} }
\decimals
\startdata
2450017.84 & ... &   ... & ... &   ... & 14.45 &   0.011 & ... &   ... & ... &   ... & ... &   ... &  VVO & H05 \\
2450021.75 & ... &   ... & ... &   ... & 14.52 &   0.024 & ... &   ... & ... &   ... & ... &   ... &  VVO & H05 \\
...&...&...&...&...&...&...&...&...&...&...&...&...&...&...\\
2460788.6308 & 15.361 & 0.009 & ... & ... & 13.970 & 0.006 & ... & ... & ... & ... & ... & ... & LOWL & H25 \\
2460789.6334 & 15.382 & 0.009 & ... & ... & 14.000 & 0.006 & ... & ... & ... & ... & ... & ... & LOWL & H25 \\
 \\
\enddata
\label{tab:photometry}
\tablecomments{
A portion of the machine-readable table, which includes all the \textit{VRIJHK} data from 1995 to 2025 as well as information about the observatories and corresponding published papers (\citet{Hamilton05}; \citet{Kusakabe05}; \citet{Herbst10}; \citet{WindemuthHerbst14}; \citet{Arulanantham17}; \citet{Aronow18}; \citet{GarciaSoto20}). The machine-readable version represents the missing magnitudes and errors with -99.999 and -9.999 (instead of ellipsis), respectively.} 
\vspace{-9mm}
\end{deluxetable*} 

\section{Results}
\begin{figure*}[htbp]
\plotone{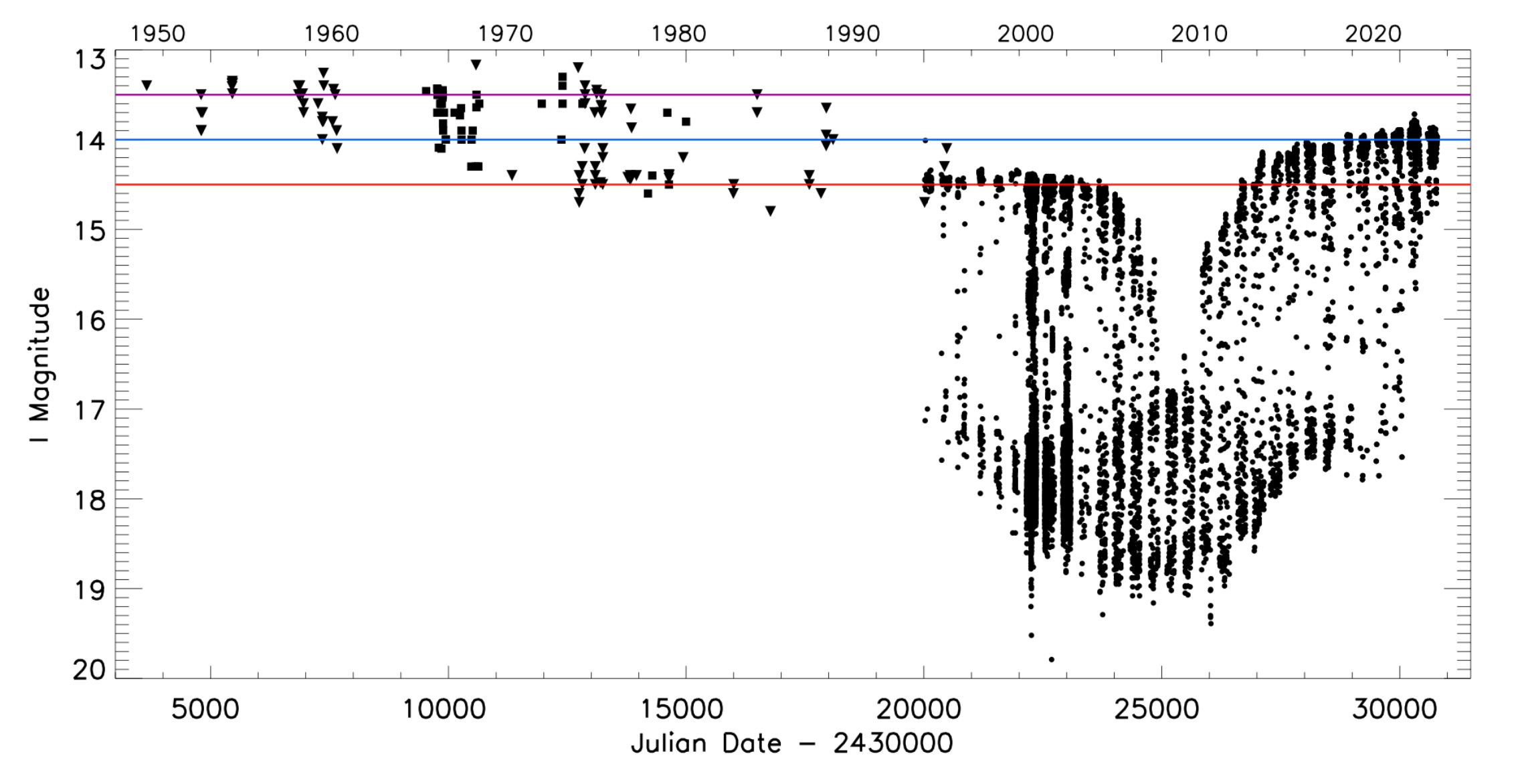}
\caption{Seventy five years of $I$ band data. The last grouping of data at the right end of the plot represent the 2024/2025 observing season. Data provided by \citet{JohnsonWinn04} (filled square) and \citet{Johnson05} (filled upside down triangle) are based on photometric measurements from photographic plates that have been transformed to the Cousins $I$ band. The purple line at $I$ = 13.5 mag represents the presumed combined magnitude of both stars A and B. The blue line at $I$ = 14.0 mag represents the magnitude of star B alone, while the red line at $I$ = 14.5 mag represents the magnitude of star A alone. See the discussion in Section~\ref{sec:appmags} regarding the $I$ magnitudes.}
\label{fig:allyears}
\end{figure*}

\begin{figure*}[htbp]
\plotone{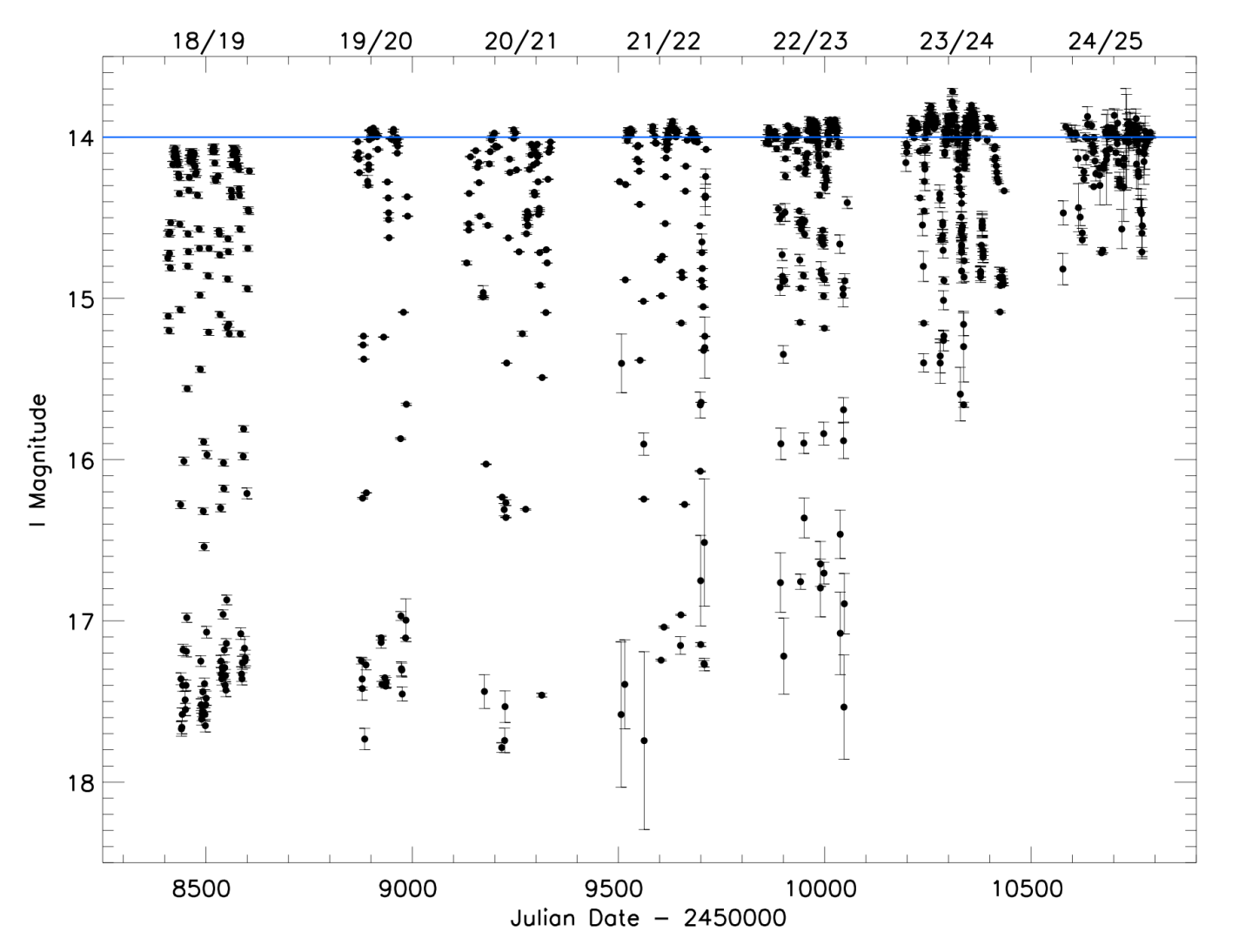}
\caption{Observations made in the $I$ band between 2018 October and 2025 April. These results are consistent with those reported in the Sloan $r$ and $g$ bands by \citet{Lamitina25}, their Figure 3. The horizontal blue line at $I$ = 14.0 mag represents the adopted magnitude of star B. The system appears to increase in brightness each season both at minimum and maximum light. Season 2024/2025 shows a slight decrease in maximum brightness relative to the 2023/2024 campaign} suggesting that the increase is not monotonic, and that stochastic influences, such as starspots, may be important. Due to the continued decrease in eclipse depths, we suggest that the era of large-amplitude photometric variability for KH 15D has ended.
\label{fig:last7}
\end{figure*}

The $I$-band light curve of KH 15D from 1951-2025 is shown in Figure~\ref{fig:allyears} and can be interpreted in light of our illustration of the system throughout the years shown in Figure~\ref{fig:schematic}. We have included photographic data available in the literature that were transformed to the Cousins $I$ band. These data are identified with different symbols and roughly cover the years from 1951 to 1997 and come from \citet{JohnsonWinn04} (filled squares) and \citet{Johnson05} (filled upside down triangles). We note that \citet{Maffei05} also produced photometric data from a subset of the same photographic plates that were used by \citet{JohnsonWinn04} and \citet{Johnson05}. Since a different process was applied to transform them to the Cousins $I$ band (see \citet{Maffei05}, \citet{JohnsonWinn04}, and \citet{Johnson05}), we choose not to include them in this figure (see \citet{Winn06} for discussion). The $I$ band magnitudes from the seven seasons of observations reported here are shown in Figure~\ref{fig:last7}. It is clear that the era of large amplitude photometric variability for KH 15D has ended, and will probably not begin again for hundreds of years. 

\subsection{Refining the Photometric Period}
\label{sec:photperiod}
\citet{KearnsHerbst98} first noted the recurrence of deep minima ($\sim$ 3 magnitudes) every 48-49 days based on the observations made over two seasons, starting in 1995 (Epoch 2 in Figure~\ref{fig:schematic}). At that time, the system exhibited a return to maximum brightness near the center of each eclipse, puzzling behavior for any type of eclipsing binary or for T Tauri stars in general, whose typical variations in brightness range from 0.1 - 1 magnitudes. This ``central reversal" of the eclipse remained an identifiable, characteristic feature of the light curve over the years. In the occulting disk model, the central reversal represents the point in the orbit where one of the stars, either A or B, comes closest to the edge of the disk during periastron passage \citep{Winn06, Poon21}. If we consider the projected ellipse on the sky (see Figure~\ref{fig:schematic}), phase 0 represents the time of maximum separation of the stars near periastron. We use this point, and the corresponding central reversal observed in the light curve as a fiducial point to define an identifiable phase zero for the photometric period. We caution the reader that this ``photometric period" may differ from the orbital period of the system, since it depends not only on the location of the stars in their orbit but also on the location and properties of the occulting screen relative to the orbit. Nonetheless, the photometric period should serve as a good guide to the orbital period and may be identical to it within the errors of both determinations.

As of now, KH 15D has been analyzed only as a single-lined spectroscopic binary \citep{Johnson04} with the periastron passage of the orbit unsampled due to its high level of obscuration in the early 2000's. When the full double-lined orbit is obtained, it should be possible to clarify the relationship between the photometric phase discussed here and the orbital configuration. In this paper we deal only with the photometric data and derive, in particular, the photometric period.  

The central reversal feature decreased in amplitude as the precession of the disk slowly occulted the orbit of star A (1995 - 2009; see Figure~\ref{fig:schematic}, Epoch 2). Focusing on the central reversal in brightness at mid-eclipse, \citet{Herbst02} established a photometric period of 48.36 days based on the first six seasons of observations. \citet{Hamilton05} reported a highly significant peak at 48.367 days in the periodogram of the photometric data collected between 1995 and 2004, which they rounded to 48.37 days. Recently, \citet{Lamitina25} confirmed a 48.36 day period based on Sloan $r$ and $g$ band data collected through the Zwicky Transient Facility (ZTF; \citet{Bellm19}) from 2017-2024 as the orbit of star B was revealed.   

\begin{figure*}[htbp]
\plotone{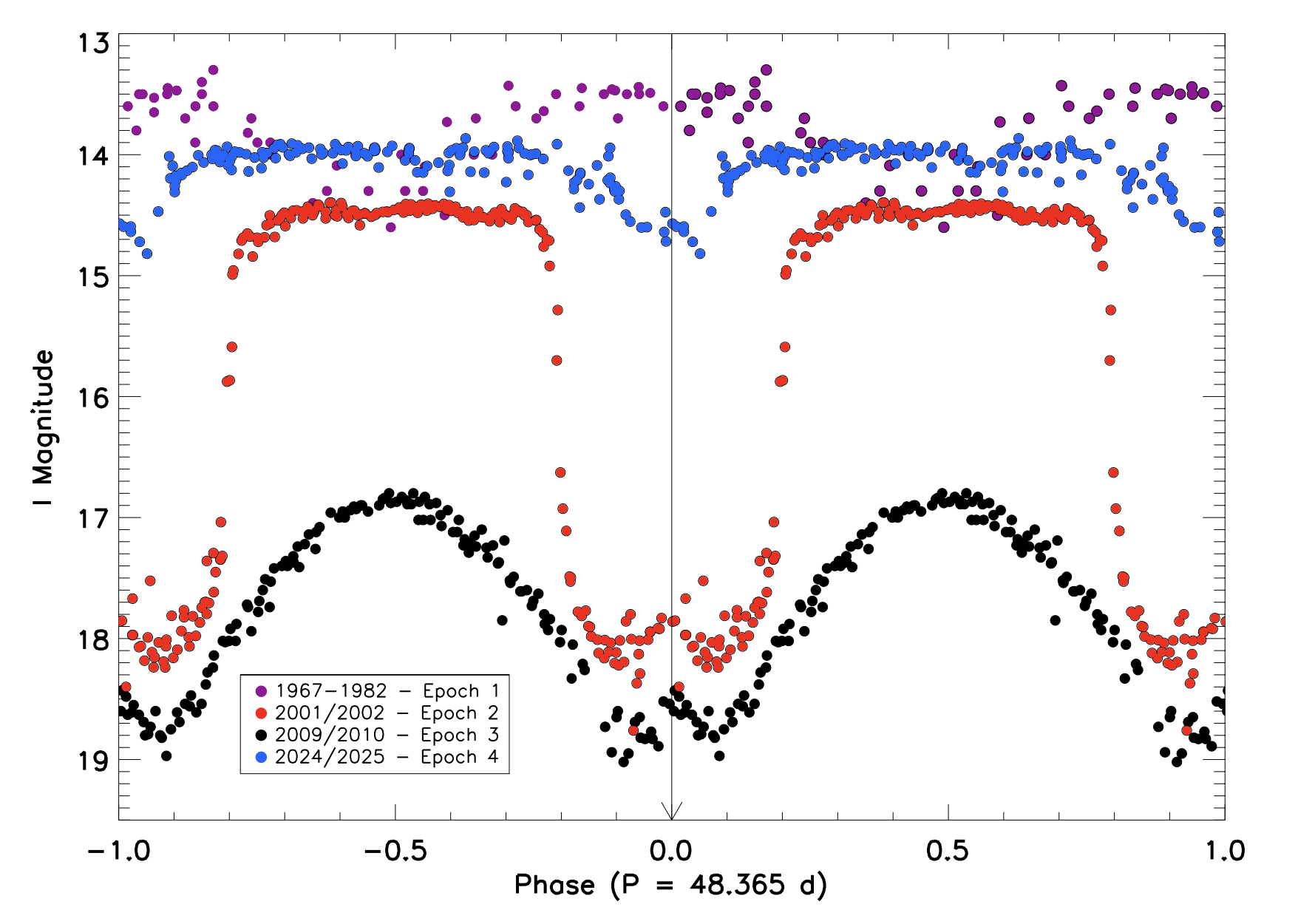}
\caption{Shown in this figure are historical light curves from 1967-1982 (purple dots) representing light coming from both stars A and B (Epoch 1 in Figure~\ref{fig:schematic}), the light curve from 2009/2010 (black dots) when  both stars were completely obscured (Epoch 3 in Figure~\ref{fig:schematic})} , the light curve from the initial international campaign conducted in 2001/2002 (red dots) when only star A was visible (Epoch 2 in Figure~\ref{fig:schematic}), and the light curve obtained in the most recent observing season representing 2024/2025 (blue dots; Epoch 4 in Figure~\ref{fig:schematic})). In this last case, we are seeing mostly star B. The alignment of the central reversals in the bottom three light curves with the central maximum in the top curve defines the photometric period, 48.365 d, as described in the text.
\label{fig:fourepochs}
\end{figure*}

To further refine the photometric period, we examined photometric data that was  representative of each epoch of light curve variability described in Section~\ref{sec:intro}. In Figure~\ref{fig:fourepochs}, Epoch 1 is represented by the historical light curve (purple dots) when the combined light from both stars is dominant (years 1967-1982, see Figure~\ref{fig:schematic}, \citet{JohnsonWinn04}; \citet{Johnson05}). At this time, star B is the star that dips behind the \textit{approaching} ring. As star B is occulted, star A remains visible and is the main contributor to the light from the system. As such, we expect the magnitude of the system to be that of star A's while B is obscured. We choose the light curve from 2001-2002 (red dots, \citet{Hamilton05}) to represent Epoch 2 identified in Figure~\ref{fig:schematic}. During this epoch, star B's orbit is fully obscured by the circumbinary ring and star A dominates the light of the system. These two epochs (purple and red data points) align at approximately $I \sim$ 14.5 mag, which is roughly the brightness of star A, supporting the model interpretation presented in Figure~\ref{fig:schematic}.

In 2009/2010 (Epoch 3 in Figure~\ref{fig:schematic}, \citet{Herbst10}), the orbits of both stars are occulted by the circumbinary ring, greatly diminishing the light received from the system (black dots). A substantial variation in brightness still occurs. This represents the faint boundary of the light curve and is interpreted to represent the scattered light component of the system \citep{Herbst10}. Finally, we choose the latest season of observations (blue dots; 2024/2025) to represent Epoch 4. Figure~\ref{fig:schematic} shows that during this epoch, eclipses occur when star B again dips behind the ring. In this epoch, however, star A's motion at periastron brings it above the \textit{trailing} edge of the ring, leaving it as the major contributor to the brightness of the system while star B is obscured. The minimum brightness exhibited by the blue dots aligns well with the out-of-eclipse red dots from 2001/2002 that define the brightness of star A. 

\begin{figure*}[htbp]
\plotone{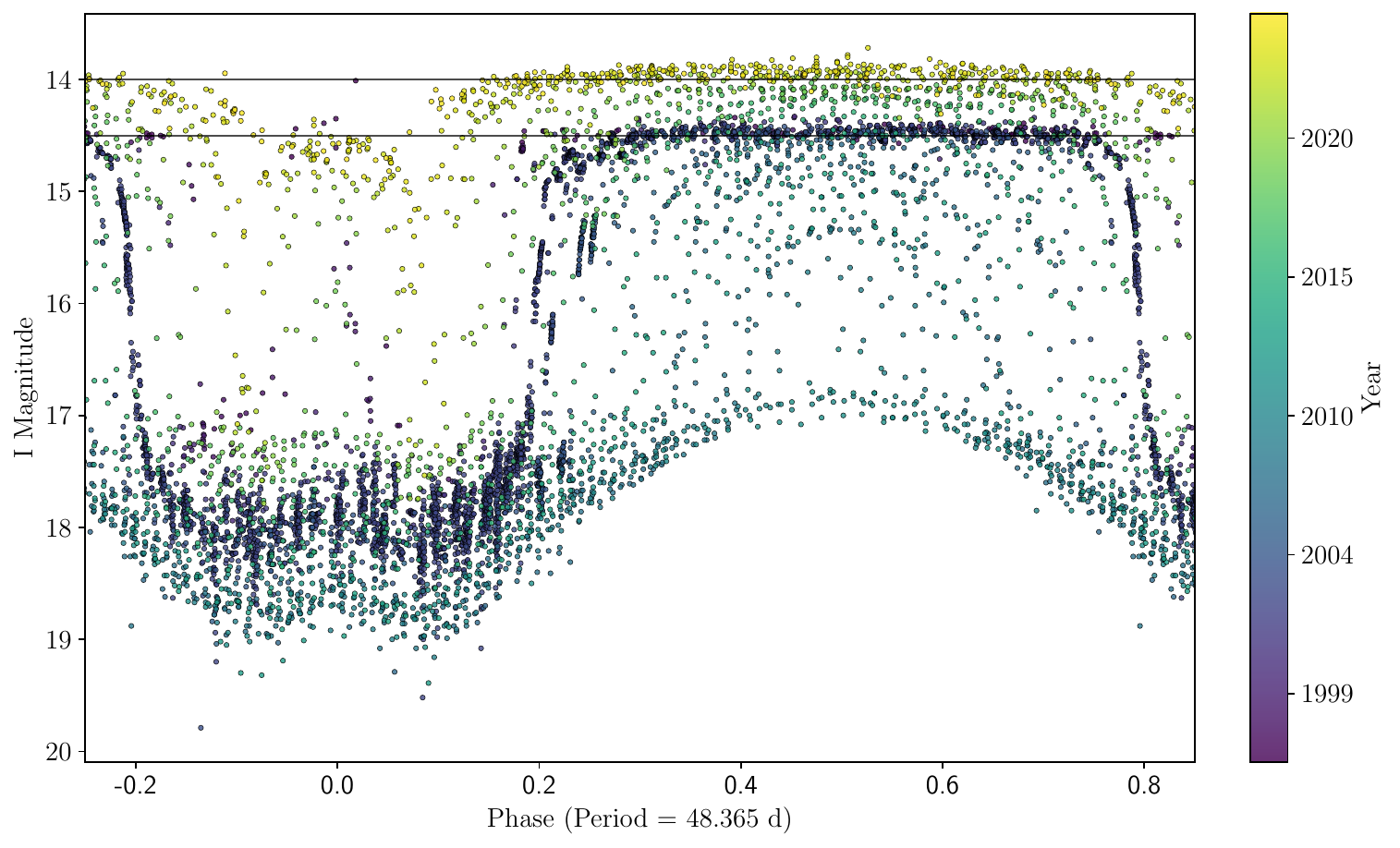}
\caption{All available photometric $I$ band data based on CCD observations listed in Table~\ref{tab:photometry} phased with the adopted period of $P$ = 48.365 d. Light curves from the early 2000s (shown mostly in purple) show sharper drops and recoveries in brightness, while the most recent observations (yellow) have produced more rounded light curves. This indicates that the leading edge has a sharper transition in optical depth than the trailing edge. The solid lines represent the adopted magnitudes of star A ($I$ = 14.5 mag) and star B ($I$ = 14.0 mag). }
\label{fig:alldataphased}
\end{figure*}

We phased the data from these epochs using periods ranging from 48.36 to 48.37 days and assessed the alignment of the ``central reversal" in 2001/2002, 2009/2010, and 2024/2025 by eye. We find that a photometric period of $P$ = 48.365$\pm0.005$ days (estimated error) centers the reversals seen at mid-eclipse (phase = 0.0 in Figure~\ref{fig:fourepochs}) and evenly splits the eclipses into two halves. All available CCD $I$ band photometry of KH 15D (Table~\ref{tab:photometry}) is shown in Figure~\ref{fig:alldataphased}, phased with a period of 48.365 days. The ephemeris is given by 

\begin{equation}
    \label{eq:ephemeris}
    JD_{0} = 24560719.625 + 48.365 E,
\end{equation} where JD$_{0}$ is the Julian Date associated with mid-eclipse, or phase 0, and E is the epoch (or cycle) number. The Julian Date 2460719.625 represents a phase of 0.0 that occurred on UT 2025 February 13.

In Figure~\ref{fig:alldataphased}, darker symbols show eclipses of star A by the leading edge of the screen during the late 1990's and early 2000's. They are quite steep and well-modeled by a ``knife-edge'' cutting across the stellar surface \citep{Herbst02}, i.e., an abrupt transition from optical depth zero to optical depth infinity. We also note how stable the light was out of eclipse during these times. There is only a small variation of order 0.1 mag mostly caused by the rotation of star A with a period of 9.6 days \citep{Hamilton05}. The more recent data, shown by lighter colored symbols, records eclipses of star B by the trailing edge of the screen. These eclipses are clearly different in shape, appearing more rounded and gradual in nature, which led \citet{GarciaSoto20} to argue that the trailing edge is not so much of a knife edge as the leading edge, but ``fuzzier''. Also note that out of eclipse, the system light is not as flat as it was in the 1990-2000's and has generally continued to rise in brightness, although not monotonically. This suggests that the occulting screen may not be as optically thick as it was and that light from both stars is contributing to the system light. We return to this point when we discuss the colors in Section~\ref{sec:vmi}. First we discuss our estimates for the unocculted brightness of stars A and B.

\subsection{The Unocculted Apparent Brightness of Stars A and B }
\label{sec:appmags}
Included in Figure~\ref{fig:allyears}, Figure~\ref{fig:last7}, and Figure~\ref{fig:alldataphased} are lines that represent the system brightness when both stars A and B are presumed to be fully visible ($I$ = 13.5 mag), when star B is presumed to be fully visible ($I$ = 14.0 mag), and when star A is presumed to be fully visible ($I$ = 14.5 mag). A brief discussion of how we come to these values is given below. First we note, however, that as spotted, rotating and possibly still accreting T Tauri stars, variability in $I$ at a level of $\sim$0.1 mag or more is expected for both stars. In fact, a rotation period of 9.6 days for star A was determined by \citet{Hamilton05} based on its periodic variations with an amplitude of 0.08 mag.  

We have also not yet had the opportunity to observe the system totally unobscured in the modern CCD era. According to models of the system \citep{Winn06,Poon21} the disk occultation began around 1960 and the models account for the historic light curves with some accuracy. By applying the model, we can approximate each star's magnitude by examining current and historic light curves. We start with the epoch when the light of both stars contribute to the bright state, namely, the period between 1967 and 1982 (Epoch 1 in Figure~\ref{fig:schematic}). The work of \citet{JohnsonWinn04} determined that the mean $I$ magnitude for the bright state of the system at this time is $I$ = 13.57 mag $\pm0.03$. 

Between 1996 and approximately 2008 (Epoch 2 in Figure~\ref{fig:schematic}), the system's out-of-eclipse brightness is modeled to result from star A alone, as the inner circumbinary ring edge occults all of star B's orbit. \citet{Hamilton05}, \citet{Herbst10}, and \citet{Aronow18} examined the photometric $I$-band data during that time period in different ways to determine the $I$ magnitude of star A. Here, we adopt a value of $I$ = 14.5 mag $\pm0.05$. This is consistent with the other estimates and includes a more generous error bar, which allows provision for the systematic error induced by the known variability of T Tauri stars.

The first estimate of the brightness of star B came from a single point obtained in 1995, which occurred very close to mid-eclipse (see \citet{Hamilton05} for details) when star A was behind the circumbinary ring. This point was measured to be $I$ = 14.01 mag $\pm$0.01 \citep{KearnsHerbst98}. The next opportunity to observe star B came in 2012 \citep{Capelo12} when it first began to emerge from the other side (the outer edge) of the occulting circumbinary ring (see Figure~\ref{fig:schematic} for reference). From the data collected in 2017-2018, \citet{GarciaSoto20} determined that the mean out-of-eclipse magnitude (brightness between phases 0.4 and 0.6) during that season was $I$ = 14.08 mag $\pm$0.01 and adopted that magnitude as the unobscured brightness for star B.

It is difficult to say whether we are seeing star B completely unocculted, or if there is still some intervening material affecting the observed brightness. As precession carries the binary to a more transparent line of sight it should be possible to establish the out-of-eclipse brightness and colors of the components more securely. In the meantime, we adopt a value of $I$ = 14.0 mag $\pm$0.05 to represent the brightness of star B, which is supported by the weighted average brightness as measured out-of-eclipse between 2016 and 2025. Using this value of $I$ for star B along with the mean $I$ magnitude for star A, we find the combined brightness of these two stars in the $I$ band to be 13.47 mag, which we round to $I$ = 13.5 mag $\pm0.05$. 

\subsection{The V-I Color Variations}
\label{sec:vmi}
\begin{figure*}[htbp]
\plotone{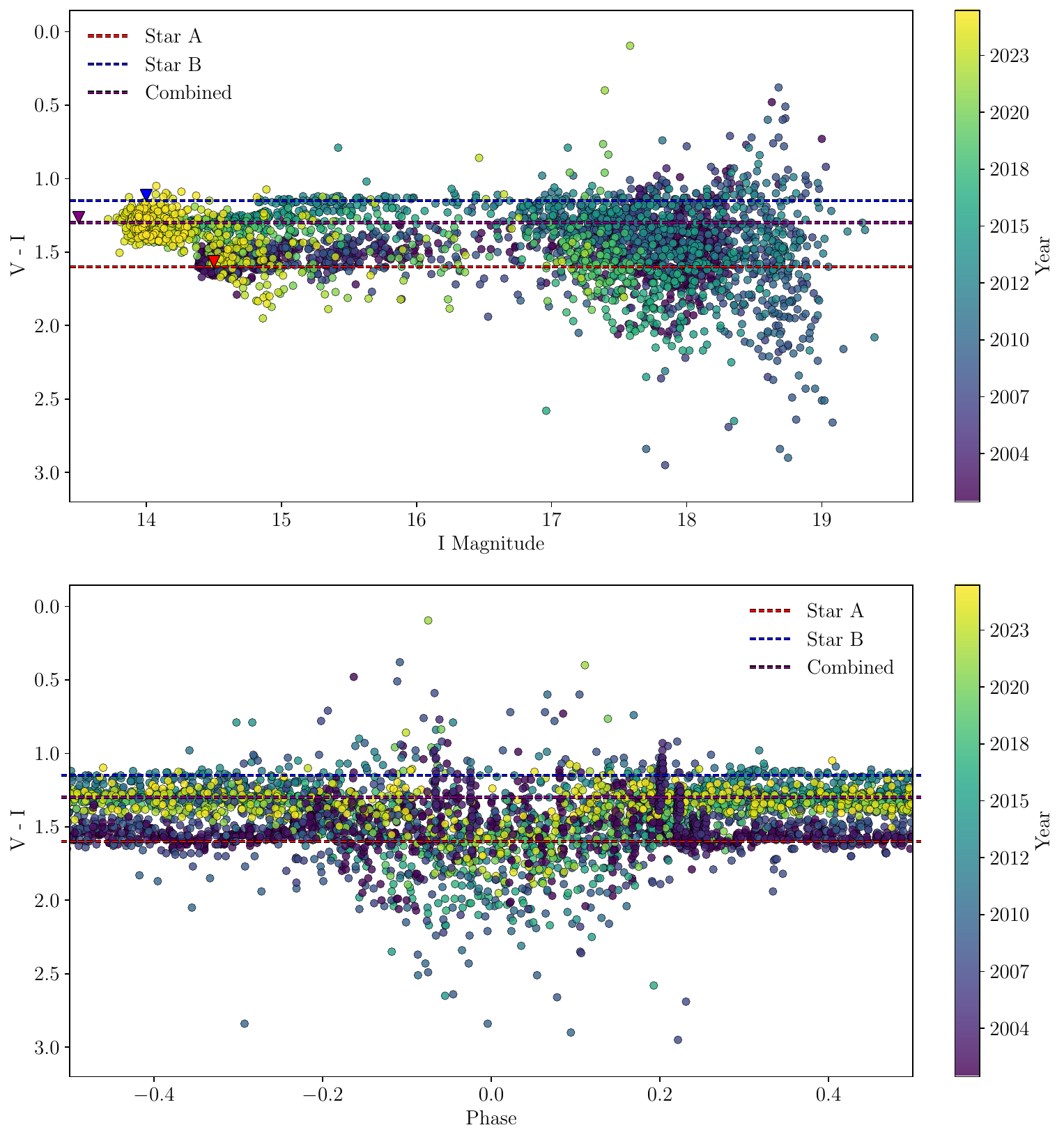}
\caption{$V-I$ color vs. $I$ magnitude (top) and vs. photometric phase (bottom) for all available data going back to 1995. Bluer colors are toward the top of the plots. The adopted colors are represented by dashed lines and the adopted magnitudes for star A, star B, and stars A+B are marked by inverted triangles.}
\label{fig:allcolorplot}
\end{figure*}

Figure~\ref{fig:allcolorplot} shows all of the available $V-I$ color measurements of the KH 15D system going back to 1995 plotted vs $I$ magnitude and photometric phase. There is obviously a good deal of variability in color, some of which depends on both overall system brightness and phase. The bottom panel shows that outside of eclipse (phases $>$ $|$0.4$|$) the color was quite stable, averaging $V-I$ = 1.55 mag in the early 2000's (darker colored symbols), when star A dominated the light at those phases, and $V-I$ = 1.15 mag around 2010 when star B took over (see \citet{WindemuthHerbst14} for discussion of the spectral types and $V-I$ color for each star). We identify those values as the intrinsic colors of each star, respectively (see Table~\ref{tab:abmags}), while reminding the reader that T Tauri stars can often vary by up to 0.1 mag in $V-I$ color due to changes in spot coverage. The predicted value of the system color if neither star were occulted, assuming $I$ = 14.5 mag for star A and $I$ = 14.0 mag for star B, is $V-I$ = 1.29 mag, which we round to 1.30 mag, as shown on both figures. In recent years (lighter colored symbols) the out-of-eclipse colors indicate that we are, in fact, seeing a mixture of light from both stars. 

\begin{deluxetable}{lccc}[htbp]
\label{abmags}
\tablecolumns{4}
\tablecaption{Adopted Unoccluded Brightness and Colors of Stars A, B, and A+B}
\tablehead
{ 
\colhead{Star} &  \colhead{Apparent $I$ Magnitude} & \colhead{$V-I$ Color} & \colhead{Reference} }
\decimals
\startdata
A  & $14.47\pm0.03$ & {$1.55\pm0.05$} & \citet{Hamilton05} \\ 
A  & $14.456 \pm0.007$ & $1.575\pm0.004$ & \citet{Herbst10} \\
A  & $14.489\pm0.001$ & $1.550\pm0.003${$^a$} & \citet{Aronow18} \\ 
A  & $14.5\pm0.05$ & {$1.55\pm0.05$} & this paper \\
B  & $14.19\pm0.06${$^b$} & $1.15\pm0.02${$^c$} & \citet{WindemuthHerbst14} \\
B  & $14.198\pm0.009$ & $1.311\pm0.009${$^d$} & \citet{Aronow18} \\ 
B  & $14.08\pm0.01$ & $1.30\pm0.01${$^e$} & \citet{GarciaSoto20} \\ 
B  & $14.0\pm0.05${$^f$} & $1.15\pm0.05${$^g$} & this paper \\
A + B  & $13.57\pm0.03$ & {...} & \citet{JohnsonWinn04} \\ 
A + B  & $13.5\pm0.05${$^h$} & {$1.30\pm0.05$} & this paper \\
\enddata
\tablecomments{
\newline
$^a$ This value is calculated using the apparent magnitudes for star A reported in Table 3 of \citet{Aronow18}.
\newline
$^b$ Predicted value based on combined light measurement of $I$ = 13.57 mag for star A and B from \citet{JohnsonWinn04} and the measured $I$ value of star A from \citet{Hamilton05}.
\newline
$^c$ Note that this value is a median, not a mean, and comes from the year 2012-2013, when star B first reappeared beyond the trailing edge of the ring.
\newline
$^d$ This value is derived by using the reported values of star B from Table 3 of \citet{Aronow18}.
\newline
$^e$ This value is derived by using the reported values of star B from Table 6 of \citet{GarciaSoto20}.
\newline
$^f$ Determined from the weighted average brightness as measured out-of-eclipse for the years 2016-2024.
\newline
$^g$ We adopt this value from \citet{WindemuthHerbst14}.
\newline
$^h$ Predicted.
}
\label{tab:abmags}
\vspace{-9mm}
\end{deluxetable}

\begin{figure*}[htbp]
\centering
\includegraphics[scale=0.7]{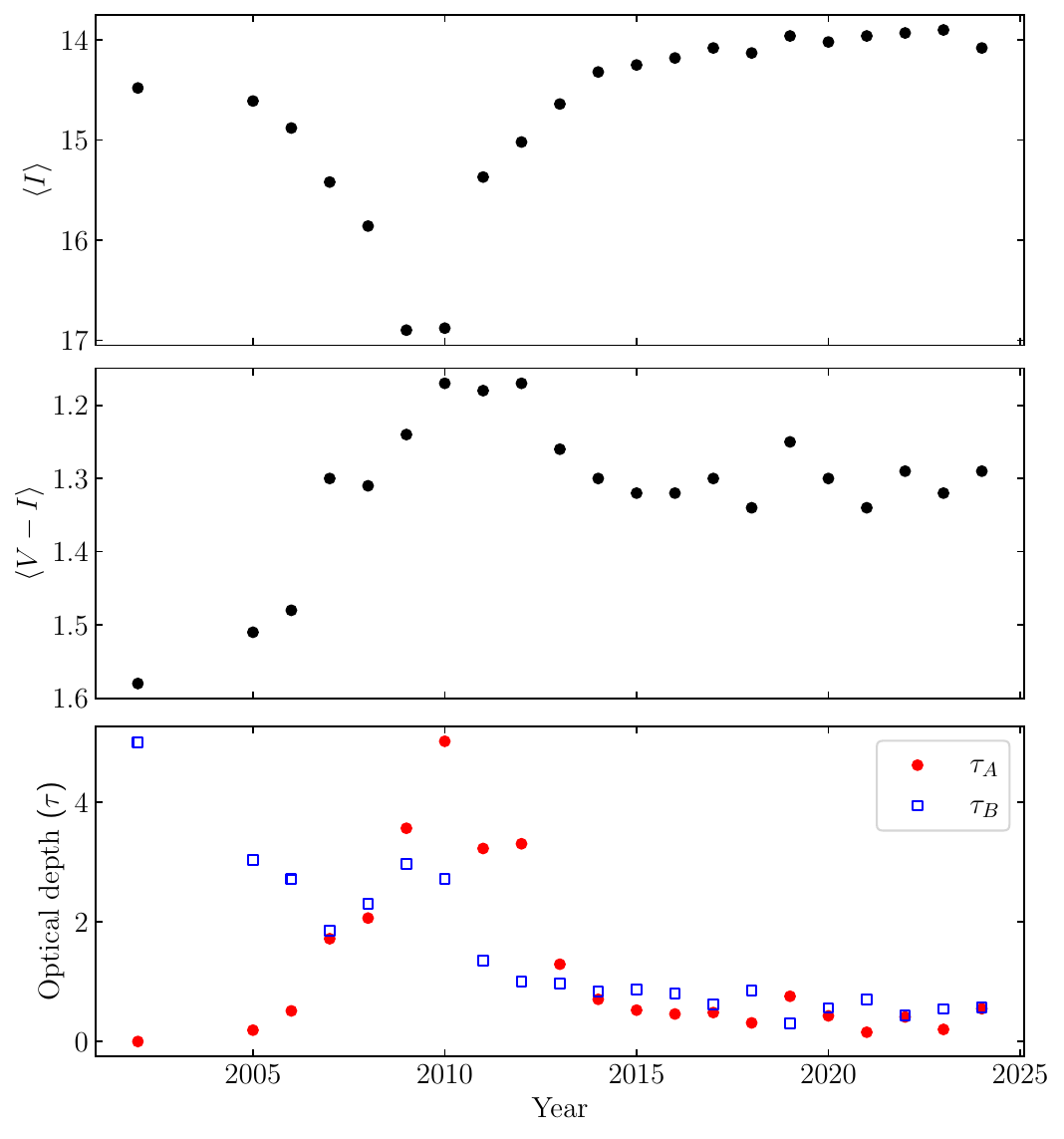}
\caption{The top two panels show the average $I$ magnitude and $V-I$ color of the system near aphelion (phases 0.4-0.6) for each year. The bottom panel shows the optical depth of the disk through which each star would need to be seen to reproduce the observed combined light magnitude and color, assuming unattenuated values of (14.5 mag, 1.55 mag) and (14.0 mag, 1.15 mag) for stars A and B, respectively, and gray extinction. Note that this is an ``effective optical depth" in the sense that it is equal to the actual optical depth \textit{only} if a single optical depth applies to the entire stellar disk. If, as appears to be the case from 2002-2010, the occulting screen is sharp and opaque then the actual optical depth may approach zero for part of the star and infinity for the other part. The system brightness and color indicates that, since about 2013 the light of both stars contributes to the system brightness at these phases. This may indicate that the trailing edge of the occulting screen is not as sharp or optically thick as its leading edge.}
\label{fig:opticaldepthaphelion}
\end{figure*}

To quantify the discussion, we show in the top two panels of Figure~\ref{fig:opticaldepthaphelion} the average values of $I$ magnitude and $V-I$ color at out-of-eclipse phases (0.4-0.6) over time. The big drop in brightness from 2002 to 2010 as star A became progressively more occulted by the highly opaque leading edge of the screen was accompanied by an increasing blueness of the system light. At that time, nothing was known about star B and the bluing of the system light was attributed to selective scattering of starlight by molecules or small grains in a halo around star A or in the disk \citep{Agol04,Winn06,SilviaAgol08}. This was supported by the fact that the polarization increased near minimum light \citep{Agol04}. Spectra taken near minimum also showed a surprising abrupt reversal in radial velocity with respect to the center of mass that was interpreted as back-scattering off the far wall of the disk \citep{Herbst08}. Following the discovery of star B, however, there is an alternative explanation for the bluing and radial velocity reversal, namely that we were actually seeing the light from star B begin to dominate the system light as star A became highly attenuated. The masses and effective temperatures of the binary components are sufficiently similar that spectral differences were not noticed (see \citet{WindemuthHerbst14} for further discussion). In this interpretation, there is no need for back-scattered light and the apparent velocity reversal is simply a transition from star A to star B as the dominant light source. That also explains the bluing of the system colors without a need to resort to selective scattering -- it is simply the growing importance of light from star B to the system brightness. Note that at this epoch there was no evidence for selective extinction (reddening) affecting the light of star B.

In the bottom panel of Figure~\ref{fig:opticaldepthaphelion} we show the optical depth through which each star would need to be viewed to account for the observed magnitude and color in each year. Note that this assumes a uniform optical depth across the star's photosphere and is, therefore, labeled as an ``effective" optical depth. In fact, the leading edge of the disk has been successfully modeled for decades as a ``knife edge", in which the rise from zero optical depth to infinite optical depth takes place on a projected length scale small compared to the radii of the stars. Between 2002 and 2010 star A progressively ``set" behind the sharp horizon of the leading edge. The system did not, however, disappear, as early models predicted because, at first, star B became progressively more visible and then star A reappeared. As can be seen in the bottom panel, we are now in a situation where the light from both stars must contribute to the system brightness near aphelion, and the effective optical depth for both has dropped to optically thin values. \citet{GarciaSoto20} argue that, based on the shape of the light curves during ingress and egress and the existence of features described as ``clumps", the trailing edge is less sharp than the leading edge and the optical depth may not reach such large values, i.e., that the disk may be better modeled as semi-transparent rather than opaque. Figure~\ref{fig:opticaldepthaphelion} shows that optical depths of 0 to 1 could explain the magnitude and colors of the system that are observed.

Returning to Figure~\ref{fig:allcolorplot}, we can see that during eclipses, near phase 0.0, the color behavior is different. In the early days (circa 2004), the system became bluer during eclipse (dark data points), probably because of the growing importance of the light of star B to the system, as discussed above. By 2012, however, the system was becoming distinctly redder during eclipse, redder even than the inferred intrinsic value of star A. \citet{WindemuthHerbst14} first noted and discussed this reddening and posited the existence of a third body whose presence could be detected by excess red light when the system was faint (see also \citet{Arulanantham17}). The relatively small number of data points bluer than the adopted value for star B are mostly found when the system is very faint and have been attributed to forward scattering by large dust grains \citep{Arulanantham16}. They may also represent times when an outburst of star B, associated with pulsed accretion near periastron, has occurred \citep{Herbst10}, or may simply be increased scatter in the data when the system is very faint. Finally, we note that in the most recent data there is a distinct trend of reddening during eclipse that sometimes extends beyond even the inferred color of star A. The simplest interpretation may be that we are observing selective extinction from relatively small grains in this part of the disk, which, if true, supports the view that the trailing edge of the occulting screen is not fully opaque. 

\begin{figure*}[htbp]
\centering
\includegraphics[scale=0.8]{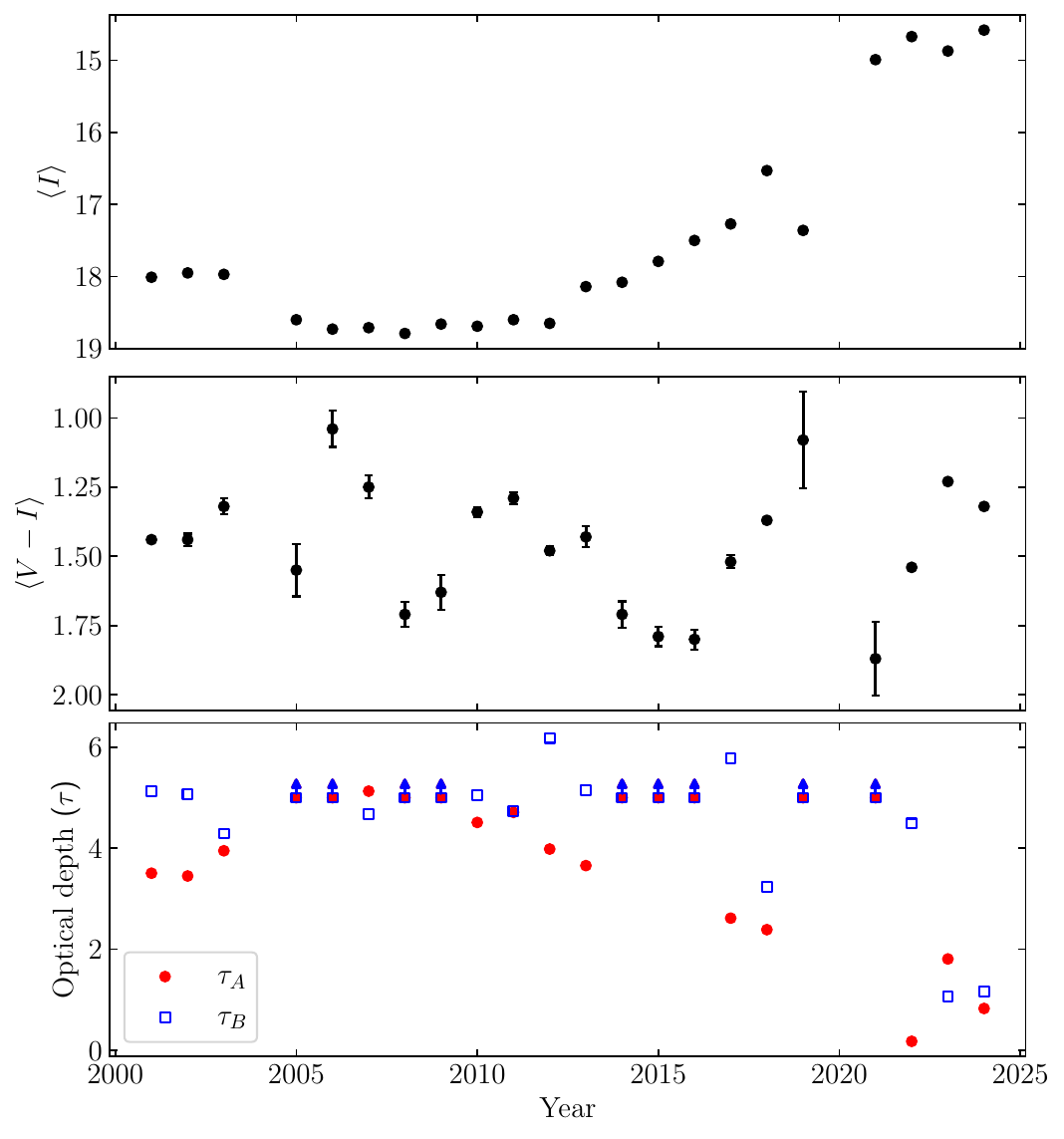}
\caption{Same as Figure~\ref{fig:opticaldepthaphelion} for photometric phases during eclipse (-0.1 to 0.1) Note that some seasons have system colors redder than our adopted color for star A so there is no solution if we assume gray extinction -- selective extinction (reddening) is required.}
\label{fig:opticaldepthperihelion}
\end{figure*}

To quantify the light and color behavior during eclipse phases we show, in Figure~\ref{fig:opticaldepthperihelion}, the average magnitude and color of the system during photometric phases $<$ $|$0.1$|$. Some of the increased scatter in $<V-I>$ can be attributed to the faintness of the object, especially in $V$, at these times. But there are a number of years when the system appears to be significantly redder than star A, precluding a solution for the optical depth that is based on an assumption of gray extinction. Reddening is required to account for these data. More recently, i.e., the last three seasons, the color may be a bit bluer, since star B is now becoming more dominant, but that does not mean that some degree of reddening is not still involved. Progress in understanding the color variation and whether it is indicative of wavelength-dependent optical depth variations in the occulting screen will require more than just $V-I$ colors. Spectra are needed to sort out the contributions of each star and establish disk transparency as a function of wavelength.

To summarize this discussion, KH 15D's color variations can be understood as primarily a result of mixing of light from stars A and B as they suffer varying levels of attenuation by the occulting disk. Between 2002 and 2012 the variations were more extreme and indicated complete attenuation of one star or the other as it passed behind the leading (inner) edge of the evidently quite opaque disk. More recently the colors suggest a mixture of starlight from both sources at most phases and some evidence for selective extinction (reddening), especially during eclipse. This supports the idea that the trailing (outer) edge is not an opaque ``knife edge" like the leading edge but a ``fuzzier", semi-transparent screen.  

\subsection{Comparison with the Poon et al. Model}
The most recent and detailed model of the KH 15D system is by \citet{Poon21} and combines the precessing warped disk proposal of \citet{ChiangMurrayClay04} with the light curve and radial velocity fitting approach of \citet{Winn06}. They include both a leading and trailing edge to the fully opaque screen and employ light ``halos" around each star as a modeling technique to account for the residual light seen even after a star is fully behind the screen \citep{Winn06, SilviaAgol08}. The ``halos" parameterize forward-scattered starlight \citep{Poon21}, which could reach us by scattering off circumbinary dust, not circumstellar dust, or it could be that the screen is not fully opaque at all lines of sight as discussed in Section~\ref{sec:vmi}.

\begin{figure*}[htbp]
\centering
\includegraphics[scale=0.6]{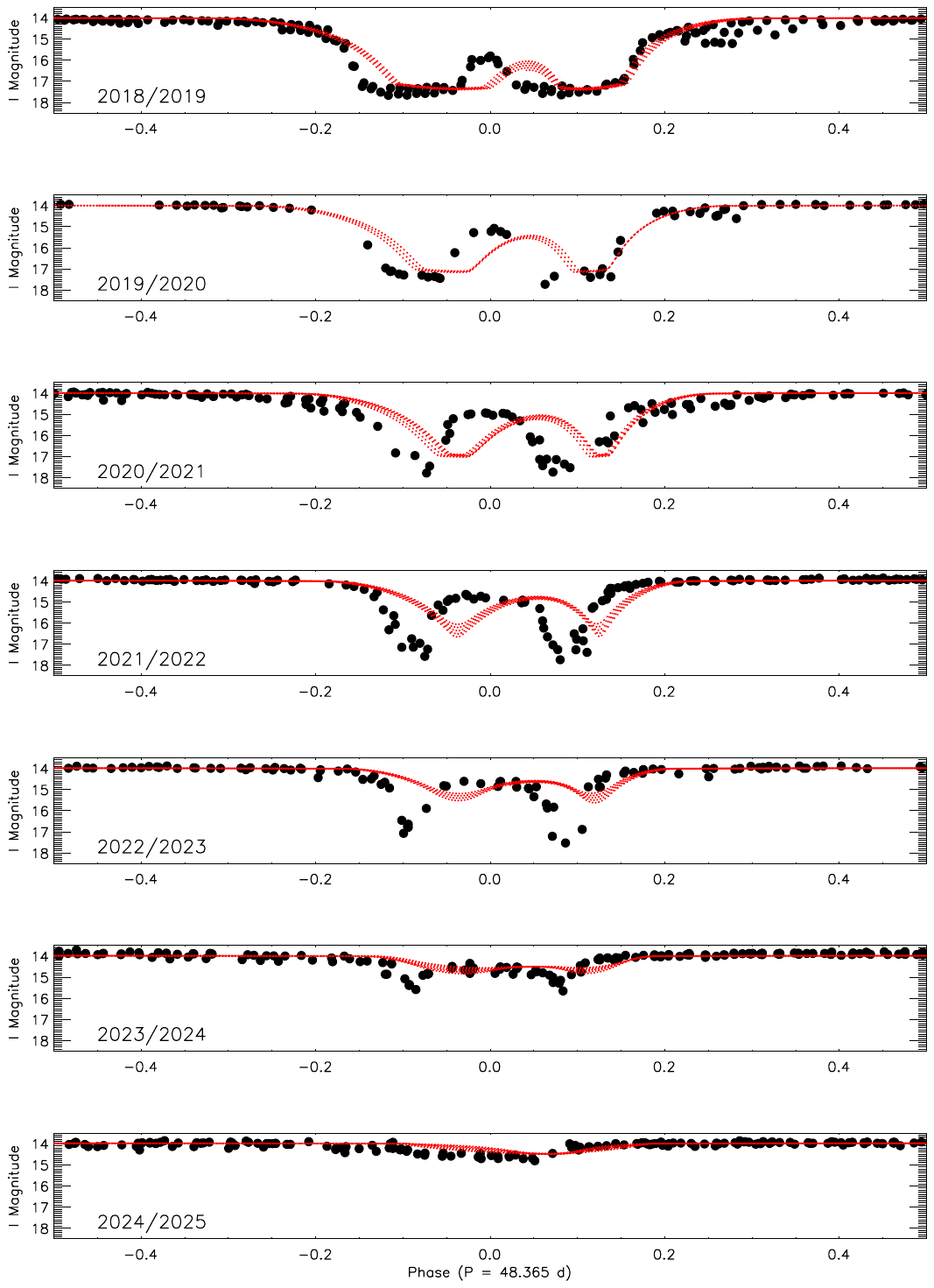}
\caption{Individual years of photometric data plotted against the model of \citet{Poon21} shown in red. Note the difference in position of the inflection points during eclipse. Note also the variations along egress from 2018/2019 through 2022/2023 and along ingress in 2018/2019 and 2020/2021. These variations are representative differences that show up in each cycle and are inferred to be caused by ``clumps" within the CB disk edge.}
\label{fig:poonmodelmatch}
\end{figure*}

The success of the current model \citep{Poon21} can be judged on Figure~\ref{fig:poonmodelmatch}, which shows our phased observations between 2018 and 2025 along with the model of \citet{Poon21}. The largest deviations occur during times of ingress and egress and during eclipse. This could be caused by slight irregularities in the actual disk edge. As \citet{GarciaSoto20} noted and \citet{Lamitina25} confirm, the trailing edge of the screen is not well-represented as a knife-edge; it is both fuzzy and clumpy in an apparently irregular manner that would be hard to model. This mismatch could also result from an evolution in the orbit, which is not modeled, such as apsidal motion or precession of the binary in the orbital plane. 

Their best model fit to the photometric and radial velocity data results in a period of $P$ = 48.377 days. We phased all of our data with a period of 48.377 days and found that the central reversal at mid-eclipse, the most notable feature that we have followed for decades, shifts in phase to the left of 0.0, in the opposite direction of where we see the model's inflection point. One possibility is that this has to do with the choice of the ``zero point" to which \citet{Poon21} have calibrated their model. The difference between the two periods is $\Delta$ 0.012 days. Thus, every cycle that is computed by the model occurs later than what would be expected based on our current ephemeris (see Section~\ref{sec:photperiod}). By 2020/2021, the shift in phase is approximately 0.055, which, using a period of $P$ = 48.377 days would correspond to a shift of approximately 2.661 days.

The model does not predict the depth of the minima very well in 2020-2023, nor does it account for the variable brightness of the system out-of-eclipse in some seasons (e.g., 2023/2024). As argued in Section~\ref{sec:vmi}, the screen through which we view the binary now is not opaque, but characterized by an optical depth of $\sim$ 1. The scattering ``halos" around each star that have been employed in the various models to explain the shape of the light curves once the star was fully behind the (assumed) opaque screen \citep{Winn06, SilviaAgol08, Poon21} may, in fact, not be necessary. It now seems likely that we have always viewed the stars through the screen, albeit with a much higher attenuation in the earlier days (1995-2008) due to the higher opacity of the leading (inner) edge. The totally opaque screen/scattering halo may simply not be that useful moving forward. 

\section{Implications and Future Work}
The James Webb Space Telescope (JWST) offers greater sensitivity, spectral resolution, and broad wavelength coverage in the near- and mid-IR compared to its predecessors. Studies of protoplanetary disks and the ices/gases present in them are now being explored in greater detail than ever before with the JWST Mid-InfraRed Instrument (MIRI) Medium Resolution Spectrometer (MRS). Several JWST survey programs (MINDS, \citet{Henning24}; JOYS, \citet{vanDishoeck25}; JDISCS, \citet{Arulanantham25}; JEDIce, \citet{Bergner26}) are providing robust results that enable comparison of protostellar/protoplanetary disks across evolutionary stage and orientation, as well as the opportunity to study the cosmochemistry taking place there. Edge-on disks provide the best insight into ices that play a key role in planet formation. In such systems, ice-phase molecules on refractory dust grains occult the host star and can produce absorption lines at near- and mid-IR wavelengths \citep{Bergner26}. In fact, ice absorption features of H$_{2}$O, CO$_{2}$, CO, and NH$_{3}$, have all been detected in edge-on systems \citep{Sturm23, Sturm24, Bergner26}. These observations reveal that ices are vertically extended, dynamically maintained, and chemically evolved.

The KH 15D system currently presents the astronomical community with an opportunity to study the dynamics and chemistry of a solar-like binary that sits within a precessing \textit{circumbinary} disk that is of planet-forming age. This circumbinary disk is modeled to extend from an inner edge at 1 AU to an outer edge at 5 AU. There is little to no observed mm emission at the location of KH 15D \citep{Aronow18} so it is difficult to know if there is an extended outer disk present. It may be that this compact disk has experienced efficient inward ``pebble drift", which has delivered solids to the inner 1-5 AU responsible for the variation in optical depth explored in Section~\ref{sec:vmi}.  \citet{Arulanantham17} postulated that the obscuring material may be condensates comprised of $\sim$10-100-micron sized particles. Based on the variability seen during ingress and egress from eclipse to eclipse (see Figure \ref{fig:poonmodelmatch}), the stars can be used as probes of this material, according to the model of \citet{Poon21}, for at least a few more years to come.

A near- and mid-IR spectroscopic study of the KH 15D system should enable the community to disentangle the contributions from dust, ices, and gas along the disk mid-plane. As the binary system illuminates different disk regions (disk atmosphere, mid-plane, and soot line) on timescales of days, it provides the opportunity to measure material properties across locations that are otherwise inaccessible in (near) edge-on systems. A single star can only illuminate a single line-of-sight in a disk, as has been described in detail in \citet{Sturm23} (see in particular their Figure 5). This single line-of-sight contains a complex compilation of direct and scattered light that arrives at our detector from various disk locations. Alternatively, as the stars in the KH 15D system orbit their common center of mass, they illuminate different sightlines within the surrounding disk. This predictable motion provides a unique opportunity to probe multiple lines-of-sight in a single disk allowing an investigation of how properties might vary with radius and altitude, without the complication of source versus source variations. Multiple ice species could be identified simultaneously, and their relative abundances compared across different disk regions. We expect to sample the H$_{2}$O ice line between 3-5 AU. Icy pebbles that migrate inward will sublimate after crossing the water snowline \citep{Banzatti23}. This can strongly alter the volatile abundance and C/O ratios \citep{Williams25}. Additionally, these observations could reveal signatures of disk winds and the outward transport of carbon-rich material from the soot line. Both processes have been inferred in the highly inclined, $\sim$1 Myr-old transitional system ISO-Oph 37, which bridges the embedded protostar and Class II disk phases \citep{Colmenares26}.

Given that KH 15D is representative of an older, more evolved population that is not yet in its debris disk stage, near- and mid-IR spectroscopic observations of it should provide a unique opportunity to probe a critical evolutionary interval, constraining the timescales over which key molecular reservoirs evolve and help to determine whether binarity accelerates or fundamentally alters chemical pathways. This is a call to obtain spectroscopic and continued photometric observations of KH 15D as the current configuration likely probes the outer $\sim$3-5 AU of the CB disk where icy dust grains reside and icy planet formation may be in progress. 

\section{Conclusions}
We have shown that the photometrically derived period of $P$ = 48.365 days accurately accounts for the observed brightness changes of the KH 15D system over seventy five years. By adopting $I$ magnitudes that represent star A and star B ($I$ = 14.5 mag and $I$ = 14.0 mag, respectively), we show that the combined light of the system in $I$ can accurately match the brightness of the system ($I$ = 13.5 mag) when both stars are believed to be unocculted. We posit that some light from both stars has been shining through the trailing edge of the disk since star B began to emerge from behind the optically thick part of the leading edge. We show that by adopting colors for star A ($V-I$ = 1.55 mag) and B ($V-I$ = 1.15 mag), the current color of the system ($V-I\sim$1.3 mag) can be explained by a combination of starlight from each star, with occasional selective reddening. If this interpretation is correct, the scattering halos employed in the models may no longer be necessary or useful moving forward. It will be very important to continue to monitor the system photometrically to assess the brightness and colors of the stars/system to sort out some of the issues discussed here. It is possible that the stars may continue to be obscured by the circumbinary disk for decades to come, providing observers with an excellent opportunity to use the binary as a moving probe behind a possibly planet-forming portion of a circumbinary disk. This would permit studies of the physical and chemical nature of the disk if variable absorption features in the stellar spectra can be found and monitored. Space-based near-infrared spectral monitoring would be highly desirable (e.g., JWST/MIRI-MRS).

\begin{acknowledgements}
We would like to thank the anonymous referee for providing suggestions, which greatly improved the quality of this paper. We would like to thank all the contributors of observational data for their efforts towards the success of the HOYS project. The authors from Dickinson College would like to acknowledge and thank the administration for supporting Dickinson's partnership in the National Undergraduate Research Observatory (NURO) through generous funding in the Physics and Astronomy yearly departmental budget. C.M.H. acknowledges support for herself and A.B.C. from the Student-Faculty Research Funds. C.M.H. thanks Wesleyan University for supporting A.J.P. during the pandemic. This paper is dedicated to those of you who supported C.M.H. through the end of this project. You know who you are.
\end{acknowledgements}


\end{document}